\documentclass[aps,twocolumn,amsmath,amssymb,superscriptaddress]{revtex4-2}
\pdfoutput=1
\usepackage[colorlinks=true,linkcolor=blue,citecolor=blue,urlcolor=blue]{hyperref}

\usepackage{amsmath}
\usepackage{graphicx}
\usepackage{graphics}
\usepackage{bm}
\usepackage[usenames]{color}
\usepackage{colordvi}
\usepackage{color}
\usepackage{units}
\usepackage{bbm}
\usepackage{booktabs}
\usepackage{array}
\usepackage{placeins}
\usepackage{epsfig}
\usepackage{changes}
\usepackage{braket}
\usepackage{tabularx}
\newcolumntype{L}[1]{>{\raggedright\arraybackslash}p{#1}} 
\newcolumntype{C}[1]{>{\centering\arraybackslash}p{#1}} 
\newcolumntype{R}[1]{>{\raggedleft\arraybackslash}p{#1}} 
\usepackage[colorlinks=true,linkcolor=blue,citecolor=blue,urlcolor=blue]{hyperref}
\newcommand{\be}{\begin{equation}}
\newcommand{\ee}{\end{equation}}
\newcommand{\beqn}{\begin{eqnarray}}
\newcommand{\eeqn}{\end{eqnarray}}

\definecolor{mymagenta}{rgb}{1.0,0.0,1.0}
\definecolor{mycyan}{rgb}{0.0,1.0,1.0}
\definecolor{myyellow}{rgb}{1.0,1.0,0.0}
\definecolor{myorange}{rgb}{1.0,0.27,0.0}

\definecolor{dark-gray}{HTML}{a0a0a0}
\definecolor{dark-red}{HTML}{8b0000}
\definecolor{dark-green}{HTML}{006400}
\definecolor{dark-blue}{HTML}{00008b}
\definecolor{gold}{rgb}{1.0,0.84,0.0}
\definecolor{dark-turquoise}{HTML}{00ced1}

\bibstyle{apsrev.bib}

\begin{document}

\title{100 Glorious Years of the Ising Model}
\author{Muktish Acharyya}
\email{muktish.physics@presiuniv.ac.in}
\affiliation{Department of Physics, Presidency University, Kolkata, India} 
\author{Yurij Holovatch}
\email{hol@icmp.lviv.ua}
\affiliation{{\color{black}Yukhnovskii} Institute for Condensed Matter Physics, National Academy of Sciences of Ukraine, 79011 Lviv, Ukraine} 
\affiliation{\color{black}$\mathbb{L}^4$ Collaboration and Doctoral College for the Statistical Physics of Complex Systems,\\ {Lviv-Leipzig-Lorraine-Coventry}, {Europe}}
\affiliation{\color{black}Centre for Fluid and Complex Systems, Coventry University, {Coventry} {CV1 5FB}, {UK}}
\affiliation{\color{black}{Complexity Science Hub}, {1030} {Vienna}, {Austria}}
\author{Ferenc Igl{\'o}i}
\email{igloi.ferenc@wigner.hun-ren.hu}
\affiliation{HUN-REN Wigner Research Centre for Physics, Institute for Solid State Physics and Optics, H-1525 Budapest, P.O. Box 49, Hungary}
\affiliation{Institute of Theoretical Physics, Szeged University, H-6720 Szeged, Hungary}
\date{\today}

\begin{abstract}
This is an editorial article based on the reseaches on the Ising model over the last 100 years.
\end{abstract}


\maketitle

\section*{Introduction}

1925 is not only the year of the birth of quantum mechanics, but also an important milestone in statistical physics, since it was in this year that Ernst Ising published his solution to the one-dimensional case of the model  \cite{Ising_Ernst} that later bore his name. In later years, the Ising model became indispensable in the theoretical description of phase transitions and plays a central role in many questions and problems arising in interacting many-body systems.
The model itself is the result of a thorough simplification that contains the elements essential for describing ferromagnetic ordering, and which, due to its simplicity, is suitable for theoretical and mathematical investigations. The importance of the Ising model from a scientific point of view is {\color{black}also} due to its interdisciplinary applications. In addition to physics, the Ising model (and some closely related models, such as the Potts model  \cite{1982RvMP...54..235W}) is of decisive importance in chemistry, biology,  {\color{black} and other
	disciplines far beyond natural sciences.  It is in place to cite here words from Giorgio Parisi's foreword to recently published book on the
	Ising model  \cite{Folk26}: ``In theoretical physics, few threads are as deceptively simple and
		profoundly influential as the Ising model. From its origins as
		a doctoral student’s attempt to understand ferromagnetism, it
		has grown into a universal tool for understanding order, disorder, and the emergence of complexity. It is a cornerstone
		of statistical mechanics, and a paradigm that has successfully
		bridged disciplines as diverse as biology, economics, and social
		science.''  }
	
	{\color{black} Ernst Ising (May 10, 1900 -- May 11, 1998) lived a long and eventful life. He was born in Cologne
		to the wealthy Jewish family of Gustav and Thekla Ising (ne\'e L\"owe). He did doctoral studies in the 
		university of Hamburg. It was his supervisor, Wilhelm Lenz, who suggested the topic of studies that 
		became the subject of the dissertation  \cite{isingthesis} successfully defended in 1924. 
		So, in the opinion of many, the model should be deservedly called the Lenz-Ising model or Ising-Lenz. 
		Such attempts at naming have been made, but history knows no subjunctive mood, and we have a name for 
		the model that has firmly established itself in scientific usage. It is worth noting
		here that the dissertation contained much more than what is now
		commonly referred to as the solution of one-dimensional Ising model reported in Ising's paper of 1925
		 \cite{Ising_Ernst}. An interested reader is referred to Refs.  \cite{Folk26,FolkHolovatch2022,Folk23,FolkHolo2024,Folk}
		for details. 
		
		In order to continue his career as a school teacher, Ernst Ising did additional studies 
		in Berlin university getting the civil servant position at a high school. However, very soon his plans 
		for the future were destroyed by the Nazi regime that came to power in Germany. A private school for 
		Jewish children, whom Ernst Ising not only taught but also saved from pogroms, arrest by the Gestapo, 
		emigration to Luxembourg, hard physical labor in occupied Luxembourg, attempts to help his wife and 
		son survive the war - that's what he experienced at that time. After the end of World War II, in 1947, 
		the family emigrated to the USA, and only then - 23 (!) years after defending his thesis, Ernst (and Ernest 
		after acquiring American citizenship) learned that in the meantime the ``Ising model'' had appeared in 
		the scientific tradition. Perhaps this fact, as well as his talent and passion for teaching, helped him get a professorship
		in physics at Bradley University in Peoria, Illinois. Students and colleagues remember him as an 
		unsurpassed lecturer, although he never returned to active scientific work. For those who would like 
		to get acquainted with the history of the Ising model and his life in more detail -- and the threats 
		to humanity and the problems of survival are becoming more and more relevant these days -- we recommend, 
		in addition to the sources mentioned above, a series of articles by Stephen G.  Brush  \cite{Brush67,Brush83}, Sigismund Kobe
		 \cite{Kobe97,Kobe00}, and Martin Niss  \cite{Niss05a,Niss09b,Niss11c,Niss05thesis}.
	}

\subsection*{Ising model and phase transitions}

Regarding the role played by the Ising model in the field of magnetism and phase transitions, we can mention the following milestones, without claiming to present a complete list. As already mentioned, in 1925 Ernst Ising published the solution of the one-dimensional model   \cite{Ising_Ernst}. {\color{black} 
Interestingly, the Ising model Hamiltonian in the form as it is known nowadays was firstly presented
in Wolfgang Pauli's contribution to the Solvay conference in 1930   \cite{pauli1930}.}
 In 1936, Peierls proved that applying the Ising model to a square lattice and fixing the edge spins to $+$ would result in the average magnetization of a central spin being strictly positive at sufficiently low temperatures, regardless of the size of the system  \cite{Peierls_1936}. In 1941 Kramers and Wanier have shown a duality relation for the square lattice Ising model and in this way determined the temperature of the phase transition  \cite{1941PhRv...60..252K}. In 1944 in a fundamental paper Onsager has shown the solution of the partition function of the two-dimensional Ising model at zero magnetic field  \cite{PhysRev.65.117}. The original solution of Onsager has been simplified by several authors, B. Kaufman  \cite{PhysRev.76.1232};  Schultz, Mattis and Lieb   \cite{RevModPhys.36.856}; Baxter  \cite{Baxter07}, McCoy and Wu \cite{McCoy1973TheTI}  and others. In 1952 C.N. Yang has calculated the spontaneous magnetization of the two-dimensional Ising model  \cite{1952PhRv...85..808Y}. For the three-dimensional Ising model there is no exact solution, although the critical exponents are accurately known 
through {\color{black} field-theoretical renormalization group expansions    \cite{PhysRevD.96.036016,schnetz2018,schnetz2023}, MC simulations   \cite{PhysRevB.82.174433,PhysRevB.104.014426} and - most precisely - through conformal bootstrap   \cite{bootstrap_3d}.

Generalization of the Ising model is the $Q$-state Potts model  \cite{1982RvMP...54..235W}, which is solved exactly for two-dimensions at the critical point  \cite{Baxter07}. The Ising model on the triangular lattice with three-spin interactions (Baxter-Wu model) is also exactly solved  \cite{Baxter07}. \textcolor{black}{The Ising model and some of its generalizations have exact solutions on regular hierarchical lattices (see e.g. \cite{Baxter07,Bleher89,Kotorowicz22}), while the behaviour on fractal lattices and the related problem of critical behavior in non-integer space dimension remain an interesting field of study \cite{Gefen80,Genzor16,LeGuillou90,Holovatch92,Holovatch98}.}

In the presence of quenched, i.e. time independent disorder the Ising model basically considered i) with random ferromagnetic couplings  \cite{Stinchcombe_1983,Folk03} ii) with random ferro- and antiferromagnetic couplings (Ising spin glass) \cite{RevModPhys.58.801} and iii) with random fields (random-field Ising model) \cite{BELANGER1991272}. During decades intensive analytical, field-theoretical and numerical investigations have been performed to clearify the critical properties of these models. Exact solution for the Sherrington-Kirkpatrick model \cite{PhysRevLett.35.1792} (Ising spin glas in the mean-field approximation) has been obtained by Parisi \cite{PhysRevLett.43.1754}, which theory has been extremely fruitful for many other problems dealing with fluctuations and disorder, in rather different fields (e. g. neural network models \cite{doi:10.1073/pnas.79.8.2554}, optimization problems \cite{mezard:jpa-00232897}, etc.). 

The quantum version of the model is obtained by extending the Hamiltonian with a transverse field \cite{RBStinchcombe_1973,sachdev_2011}. This system is usually studied at low temperatures in the $T=0$ limit, when there is a quantum phase transition in the ground state of the system, when the strength of the transverse field plays the role of the control parameter \cite{sachdev_2011}. In homogeneous systems the quantum phase transition in a $d$-dimensional system belongs to the same universality class as its $(d+1)$-dimensional classical counterpart \cite{RevModPhys.51.659}. For disordered quantum Ising models the phase transition is infinite disorder type \cite{IGLOI2005277,Igloi2018}.

\textcolor{black}{It would be worth mentioning here that the nonequilibrium phase transitions are widely studied in Ising ferromagnet in the end of last century. The critical slowing down and specific heat sigularity \cite{muktish1}, divergence of the fluctuations
of dynamic order parameter \cite{muktish2}, dynamical symmetry breaking by randomly varying magnetic field \cite{muktish3}, connection between dynamic transition and maximum hysteretic loss \cite{muktish4}, connection between dynamic phase transition
and the stochastic resonance \cite{muktish5} are the examples of such studies. A review \cite{muktish6} contains all such works
and related studies. In the present century, the dynamic phase transition in kinetic Ising model driven by oscillating magnetic field has been studied widely. The dynamic transition in the presence of quenched randomness \cite{erol}, the comparison of surface and bulk dynamic phase transition
 \cite{michel}, the determination of the universality class \cite{rikvold} are some of the examples. Experimental evidence of the dynamic phase transition with metamagnetic anomalies have also been 
reported \cite{berger}.}

\section*{Special issue}

This special issue of the EPJB deals with the problems associated with the Ising model, focusing on current applications of the model and presenting the most important results achieved. The articles submitted for the special issue are basically of two types. One part of them – ten studies – falls into the category of thematic reviews and aims to present the basic results of the given problem area. The other part of the articles – ten manuscripts – are so-called research articles and contain the latest research results. In the following, we present the results in detail in each article, starting with topical reviews and continuing with research papers.

\subsection*{Topical reviews}

\subsubsection*{Ernst Ising's thesis}

The historical background of the Ising model is presented in the article by \textit{Reinhard Folk} \cite{Folk}, which analyzes in detail the results of Ising's doctoral dissertation. He usually supports his conclusions with quotes from famous scientists, which he also gives in the original language for the sake of professionalism. The model itself was created as a result of a very strong simplification, the individual steps of which were formulated in the works of Richard Kirwan \cite{kirwan1797}, Pierre Curie \cite{curie1894}, Paul Langevin \cite{langevin1905magnetisme}, Pierre-Ernest Weiss \cite{weiss1907hypothese}, Wilhelm Lenz \cite{lenz1920beitrag}, Walter Schottky \cite{schottky1922drehung} and others, and its authentic explanation based on quantum mechanics became complete with the discovery of the electron spin and the recognition of the exchange interaction. The article describes Ising's original calculation procedure in the one-dimensional case, which is related to the transfer matrix method that was later used in a standard way. The article also shows that Ising discussed other generalized models in his dissertation, such as the 3-state model (now known as the Potts model \cite{1982RvMP...54..235W}) and the ladder model consisting of two interacting chains. At the end of the article, we will have a brief overview of the role of the Ising model in the field of critical phenomena, Onsager's exact solution in the two-dimensional case \cite{onsager}, the behavior in the case of long-range interactions, and the results provided by the renormalization group method.

\subsubsection*{Interface structures}

Results obtained from the exact solution to the Ising model in two dimensions have played a pivotal role in statistical mechanics for more than 80 years, the importance of which cannot be overestimated. The manuscript by \textit{Alessio Squarcini, Piotr Nowakowski, Douglas B. Abraham and Anna Maciolek}, entitled "Partition function for several Ising model interface structures" \cite{Squarcini_Nowakowski_Abraham_Maciolek} is a pedagogical account of exact results obtained for various incremental contributions to the free-energy in inhomogeneous systems when interfaces and droplets are formed near pinning boundaries. This work describes just some of the body of work pioneered by D.B. Abraham and coworkers over the last 50 years. This is probably the most accessible account of the transfer matrix formulation and associated complex analysis involved in the evaluation of the integrals arising from the exact solutions - the free-energy contributions are by far the easiest to compute, compared to say the magnetization and even energy-density correlations - but even these require extremely detailed analysis, which is presented here with great clarity and lucidity. The focus of the paper is unashamedly the methods and results of the exact calculation.

\subsubsection*{Nonadditive entropies}

\textit{Henrique Santos Lima and Constantino Tsallis} discuss the entropy of one-dimensional ferromagnetic Ising-like models with longitudinal and transverse magnetic fields \cite{Lima_Tsallis}. They consider the nonadditive entropy through the definition
\be
{\cal S}_q=k \sum_i^W p_i \ln_q \frac{1}{p_i}\;,
\ee
where $\ln_q x=\frac{x^{1-q}-1}{1-q}$  is the $q$-logarithm (with $\ln_1 x= \ln x$) , $p_i$ being the probability of the $i$-th microstate and $W$ denotes the number of microstates.  The quantum version of the nonadditive entropy is given by:
\be
{\cal S}_q=k {\rm\bf Tr}\left( \rho \ln_q \frac{1}{\rho}\right)\;,
\ee
in terms of the density operator $\rho$.
For a quantum critical
system at $T=0$ at the special value
\be
q^*=\frac{\sqrt{9+c^2}-3}{c}\;,
\ee
the ${\cal S}_q$ entropy is extensive, where $c$ denotes the central charge of the Virasoro algebra (for the critical quantum Ising chain $c=1/2$, for the $XX$ chain it is $c=1$). They also discuss the issue of divergence at the critical point. The manuscript represents a contribution to the area of expertise of the authors.

\subsubsection*{Ising spin networks}

In the paper by \textit{Bosiljka Tadic and Rasa Pirc} \cite{Tadic_Pirc}, the authors study spin reversal dynamics on the hysteresis loop in Ising spin networks consisting of triangles assembled into three different architectures and triangle-embedded interactions of a random sign. The authors show that the hysteresis loop shape and spin activity avalanches during the field-driven magnetization reversal are affected by geometric frustration, in conjunction with the topology of an assembly. The latter is connected with the preference of triangles in the growing network for sharing a single node or a single angle. In the third approach, sharing of both nodes and edges is equally probable, and the network shape is mainly governed by geometrical constraints. The interactions are restricted to pairwise or triple interactions.
Independently of the assumed strategy of network growth, the hysteresis loops exhibit step-like magnetization. The big magnetization jumps are observed in the case of strongly frustrated antiferromagentic pairwise interactions. The avalanches which characterize the Barkhausen noise follow the same probability distribution.

\subsubsection*{Hysteresis}

In their paper, \textit{Deepak Dhar and Sanjib Sabhapandit} \cite{dhar_sabhapandit} provide an overview of the field of hysteresis in magnetic models. They discuss the hysteresis loop, the dynamics of symmetry breaking, and dynamic phase transitions. In particular, they address Barkhausen noise in hysteresis loops, where the broad size distribution of magnetization jumps can be modeled by the random-field Ising model \cite{BELANGER1991272}. They discuss the size distribution of these jumps in the random-field Ising model on the Bethe lattice. They also discuss early work in this field, including the Preisach model, the scaling area of the hysteresis loop, the hysteresis loop in Ising and Ising-like models, in continuous-spin models, dynamic phase transitions, in disordered magnets, and the distribution of avalanches.

\subsubsection*{DNA unzipping}

The review article by \textit{Somendra Mohan Bhattacharjee} \cite{Bhattacharjee}, explores the force-induced unzipping of double-stranded DNA
(dsDNA). The manuscript begins with a historical overview of significant
discoveries in both molecular biology and physics post-World War II,
highlighting the parallel advancements that shaped our understanding of DNA.
Key biological breakthroughs included identifying DNA as genetic material
through experiments like Avery-MacLeod-McCarty (1944) \cite{10.1084/jem.79.2.137} and Hershey-Chase (1952),
and the elucidation of its double-helical structure by Watson and Crick in
1953 \cite{crick_watson}. Concurrently, physics saw developments such as Onsager's solution to the
2D Ising model (1944) \cite{PhysRev.65.117}.
The separation of strands by breaking hydrogen bonds is crucial for biological
processes like replication and gene expression. The manuscript then reviews the
coarse-grained theoretical frameworks for analyzing thermal melting of DNA,
emphasizing universal features independent of microscopic details. The
Poland-Scheraga model \cite{10.1063/1.1727785} describes DNA as alternating bound (helical) and unbound
(denatured loop) segments, separated by junction points. The Zimm model \cite{10.1063/1.1731411}, a
variation, assumes separated strands cannot rejoin and exhibits a first-order
transition. These models use partition functions and singularities to determine
phase transitions, with the behavior of bubbles (denatured loops) playing a
critical role. For instance, a continuous melting transition is possible if
bubbles can form within the bound state, and is characterized by a diverging
length scale  \cite{Honchar21}.
Unlike thermal melting, where base pairs disrupt throughout the molecule,
unzipping initiates at the ends and propagates along the DNA, with the applied
force stretching the separated strands. The unzipping transition is crucial
because melting temperatures are often too high for physiological conditions.
The manuscript reviews coarse-grained models used to study the DNA unzipping
transition with constant and periodic pulling forces. It discusses in detail
the exact the phase boundary, low temperature reentrance, dynamics and
hysteresis associated with DNA unzipping transition.

\subsubsection*{Sociophysics}

The  article by \textit{Pratik Mullick and Parongama Sen} presents a comprehensive and pedagogical review of the role of the Ising model and its variants in the field of sociophysics \cite{Mullick_Sen}. The review covers a wide range of applications, including binary opinion dynamics, social balance, norm breaking, segregation phenomena, and language evolution. The paper bridges statistical physics and social modeling by highlighting how the Ising framework has been adapted and generalized to capture essential features of collective human behavior. Overall, the review is of interest to researchers working in complex systems, computational modeling, and statistical physics applied to interdisciplinary contexts.

\subsubsection*{Ising spin glass}

The recent developments of Ising spin glass are briefly reviewed by \textit{Purusattam Ray} \cite{purusattam}. The quenched disorder and frustration
in the Ising model lead to spin glass phase. The theoretical understanding of the emergent glassy phase in the discrete symmetric Ising model is discussed through the well known Edward-Anderson (EA) model  and Sherington-Kirkpatrick (SK) model. The 
comprehensive phase diagram is shown  for SK model. The phase diagram of SK model with normally distributed 
spin-spin interaction  contains the Almeida-Thouless
line separating the region where replica symmetric solution is stable from the region where replica
symmetry breaking occurs. The Parisi solution of replica symmetry breaking scheme has been discussed in this context. The
critical exponents are provided . A short interesting discussion on the use of  physics of Ising spin glass in econophysics and sociophysics is also found in this article.

\subsubsection*{AI and ML in Ising model}

The use of artificial intelligence and machine learning has been discussed in an article written by \textit{V. Babu and R. Pandit} \cite{rahul}. The machine learning (both supervised and unsupervised) can be used to study the Ising 
ferromagnetic phase transitions. \textcolor{black}{The two dimensional Ising ferromagnet with next nearest neighbour interaction and the three
dimensional Lenard-Jones model for a continuum fluid (in the vicinity of liquid-gas phase boundary this can be mapped to
the Ising model), the algorithms of neural networks trained with the spin configuration are used to study the liquid-gas
phase transition.}

\subsubsection*{Coarsening kinetics in long ranged Ising model}

The main features of one dimensional long-ranged (with power law decay of spin-spin interaction)
 Ising model are briefly reviewed by \textit{Frederico Corberi, Eugenio Lippiello, Paolo Politi and Luca Smaldone}
 \cite{corberi} in the context of voter model. In the nearest neighbour voter model, a randomly chosen agent takes the
state of that for any (randomly chosen) of its nearest neighbours. In this study, this idea has been generalized. In the long-ranged voter model, in an elementary move, a randomly
chosen agent takes the state of another one, chosen at distance $r$ with probability $P \sim r^{-\alpha}$. In the $p$-voter model,
the agent takes the state of partial spin average of $p$ number of neighbouring spins. In this article, the authors show,
that this $p$-voter model belongs to the Ising universality class for $p \geq 3$.

\subsection*{Research papers}

\subsubsection*{Non-equilibrium processes}

In the paper by \textit{Ian Filippo Schönherr, Fabio Müller and Wolfhard Janke} the authors use a rather new very efficient Monte Carlo algorithm for long-range models, algebraically decaying with an exponent $\sigma =0.9$ or $\sigma =1.1$ to investigate non-equilibrium phase-ordering kinetics in the two-dimensional Ising model with non-conserved order parameter \cite{Schnherr_Muller_Janke}. They quench the system from $T = \infty$ to two quench temperatures, $T_q = 0.1 T_c $ and $T_q = 0.5 T_c$, $T_c$ being the critical temperature.
The main results obtained are that the scaling hypothesis holds for
both quench temperatures and for both values of sigma, with a
corresponding growth exponent as predicted in earlier works. The
authors also show that at the higher quench temperature, there exists
an extended scaling regime; that is, the predicted growth exponent is
valid over a larger time window.

In the paper by  \textit{Ramgopal Agrawal, Federico Corberi, Eugenio Lippiello and Sanjay Puri} the authors consider the random field Ising model (RFIM) in 1d and 2d with long range interactions (algebraically decaying with an exponent $\sigma$) and study numerically the domain growth after a rapid quench to low temperatures \cite{Agrawal_Corberi_Lippiello_Puri}. The interesting question here is whether the long-range interactions can accelerate the dynamics which otherwise, for short-range interactions, would be activated and thus logarithmically slow. In 1d they find that the asymptotic domain growth law is logarithmic for all values of the exponent $\sigma$. In 2d they find that for small disorder amplitudes and low temperature quenches maintains a transient power-law growth for domain size. When the disorder amplitude exceeds the thermal energy scale they observe an extremely slow domain growth.

The paper by \textit{Subir Das and Soumik Ghosh} concerns the numerical study of coarsening in the 2d Ising model with long-range interactions, algebraically decaying with an exponent $\sigma$ \cite{Das_Ghosh}. The order parameter in the MC simulations was conserved via the incorporation of the Kawasaki spin-exchange method. Their results show that growths follow power-laws, with the exponent which depends on the range of interactions. When the range is above a cut-off, the exponent,
for any given range, seems to change from a larger value to a smaller one, during the
evolution process.

The paper by \textit{Annalisa Fierro, Antonio Coniglio and Marco Zannetti} studies the effect of two types of boundary conditions on the dynamics of the Ising model in 1d (through analytical results) and in 2d (through MC simulations) \cite{fierro_coniglio_zanetti}. They find that as long as the time dependent correlation length $R(t)$ is smaller than the length of the system,
both boundary conditions show similar dynamics. However, when this correlation length is of the order of system size when finite size effects dominate,
periodic boundary conditions (PBC) and antiperiodic boundary conditions (APBC) show different behaviour. For quenches below $T_c$, while PBC shows
decrease in $R(t)$ and fluctuations of order parameter due to ergodicity and symmetry breaking, APBC shows increase in $R(t)$ and fluctuations as system size increases hinting towards a critical equilibrium state for APBC, even when the quench is below $T_c$ and need not be at the critical temperature $T_c$.

\subsubsection*{Critical properties}

In the paper by \textit{Nabil Hachem, Abderrahim Ezzaime and Mohammed El Bouziani} the authors study the critical behaviour of the three-dimensional semi-infinite spin-2 Ising model using Migdal-Kadanoff renormalization group method \cite{Hachem_Ezzaime_ElBouziani}. They propose a phase diagram which consists of
ordinary second order phase transition line, first order phase transition line merging at a tricritical point with the second order line. In addition to these, they have also obtained a phase
where the bulk is disordered and the surface is still ordered, where these new phases are separated by the conventional phases by second order phase transition line. As the authors also comment,
these results are in line with semi-infinite spin-1 Ising model. Hence, the same result is obtained on a new system.

In the paper by \textit{Abdelhak Boukhal, Youssef Chegrane, Nabil Hachem and Mohammed El Bouziani}, the magnetic behavior of the $CrI_3$ bilayer was investigated within the mean field approximation (MFA) using the spin-3/2 Ising model \cite{Boukhal_Chegrane_Hachem_ElBouziani}. Understanding the magnetic properties of 2D materials such as $CrI_3$ is of great importance for spintronic and magnetic memory applications. The magnetic properties of the $CrI_3$ bilayer were modeled with an Ising-type Hamiltonian and analyzed using MFA. Key findings include {\color{black} investigation of} the effects of the interlayer interaction ($J_{int}$) and crystal field (D) parameters on the critical temperature ($T_c$). {\color{black}Besides, a} second-order ferromagnetic-paramagnetic transition was observed {\color{black} and a}  first-order transition between the F1/2 and F3/2 ferromagnetic phases was detected at low temperatures. The variation of the blocking temperature ($T_B$) with the external magnetic field and crystal field was reported. 

The paper by \textit{Lourens Waldorp, Tuan Pham and Han van der Maas} explores an extension of the Ising model, having $2k+1$ different states what is termed the "general-spin Ising model" \cite{Waldorp_Pham_vanderMaas}. They analyze the model using a mean-field approximation through variational principle of the Gibbs free energy on a regular graph. The  inspiration of the authors to use such model comes from the use of statistical physics models in sociology
and psychology. The model exhibits spontaneous magnetisation, similar to
the standard Ising model and the phase transition follows the same route, just the location depends on the value of $k$.

\subsubsection*{Quantum models}

The work by \textit{Soumyaditya Das, Soumyajyoti Biswas and Bikas K. Chakrabarti} studies the quantum Sherrington-Kirkpatric (SK) model in the transverse field with the Suzuk-Kubo-de Gennes dynamics \cite{das_biswas_chakrabarti}. The authors have studied simulated annealing (SA) of the SK model with the same dynamics in the previous work (ref. [7]) and this work is an extension to quantum annealing (QA). The main finding is that quantum annealing and classical (simulated) annealing basically yield the same result. This result implies that there is no supremacy of QA over SA as far as the Suzuki-Kubo's mean-field dynamics is employed. Although this might be a negative result of QA, the comparison of QA with SA is meaningful.

The paper by \textit{Ali Aali and Farrukh Mukhamedov} deals with a Quantum Markov Chain (QMC) associated with
mixed quantum Ising-XY model on a Cayley tree of order two \cite{AlAali_Mukhamedov}. The considered
mixed model has the nearest-neighbour Ising interaction $J_I$ at odd levels of the
tree, and the nearest-neighbour XY-interaction $J_{XY}$ at even levels of the tree.
It turns out that for this mixed model there is only unique periodic QMC which is translation-invariant
one. Moreover, one can construct non translation-invariant QMC for the model.

\subsubsection*{Metastability}

The paper by \textit{Franco Bagnoli and Tommaso Matteuzzi}, studies metastability phenomena in parallel Monte Carlo implementations of the Ising model \cite{bagnoli_matteuzzi}. Specifically, the authors investigate checkerboard patterns arising in parallel Glauber dynamics, compare serial, Wolff, and parallel updates, analyze the correlation index $c(t)$ as an order parameter, and examine finite-size scaling of escape times under partial asynchronism (dilution).

\begin{acknowledgments}
The work by F.I. was supported by the National Research Fund under Grant No. K146736, and by the National Research, Development and Innovation Office of Hungary (NKFIH) within the Quantum Information National Laboratory of Hungary. 
\color{black}{Yu.H. was supported by the National Research Foundation of Ukraine Project 2023.03/0099
``Criticality of complex systems: fundamental aspects and applications''.}

\end{acknowledgments}

\bibliography{Bibliography_editorial}

\end{document}